\documentclass{nsr}

\usepackage{amsmath,graphicx,array}
\usepackage{dcolumn,soul}%

\usepackage{amsthm}
\usepackage[figuresright]{rotating}%
\usepackage{algorithm, algorithmicx, algpseudocode}
\usepackage{listings}%
\usepackage{hyperref}
\usepackage{bm}

\makeatletter
\def\uns{\ifmmode\,\else$\,$\fi}%

\makeatother
 
\jvol{XX}
\jnum{X}
\jyear{2026}
\doi{10.1093/nsr/XXXX}
\received{XX XX Year}
\revised{XX XX Year}
\accepted{XX XX Year}

\begin{document}

\dhead{RESEARCH ARTICLE}

\subhead{INFORMATION SCIENCE}

\title{WiFo-2: a generalist foundation model unifies heterogeneous wireless system design}

\author{Boxun Liu$^{1}$}
\author{Xuanyu Liu$^{1}$}
\author{Shijian Gao$^{2}$}
\author{Xuesong~Cai$^{1}$}
\author{Xiang Cheng$^{1\ast}$}
\author{Liuqing Yang $^{2,3}$}

\affil{$^{1}$State Key Laboratory of Photonics and Communications, School of Electronics, Peking University, Beijing, 100871, China}

\affil{$^{2}$Internet of Things Thrust, The Hong Kong University of Science and Technology (Guangzhou), Guangzhou, 511400, China}

\affil{$^{3}$Intelligent Transportation Thrust, The Hong Kong University of Science and Technology (Guangzhou), Guangzhou, 511400, China}

\authornote{\textbf{Corresponding authors.} Email: xiangcheng@pku.edu.cn}

\abstract[ABSTRACT]{Emerging sixth-generation wireless systems are increasingly heterogeneous, with compatibility across diverse configurations, ubiquitous coverage, and expanded functionalities.
Although deep learning has substantially benefited wireless system design, existing approaches are typically trained for specific system settings and scenarios with limited generalizability. 
Here we present WiFo-2, a space-time-frequency foundation model for unified wireless communications and sensing system design.
Pretrained on a heterogeneous dataset of 11.6 billion channel state information (CSI) points, WiFo-2 learns generalized wireless representations across scenarios, configurations, and tasks, and exhibits scaling-law behavior.
WiFo-2 achieves reliable and accurate zero-shot channel reconstruction, outperforming fully supervised task-specific models.
With only 1\% of the training samples required by supervised AI models, WiFo-2 achieves state-of-the-art performance across 9 distinct wireless tasks. 
A functional hardware prototype further demonstrates its real-world deployability and superior capability across diverse wireless tasks.
This work provides a versatile wireless design framework and advances understanding of wireless channels.}

\keywords{channel state information, foundation model, pretraining, wireless system}

\maketitle

\section{INTRODUCTION}
Wireless technology underpins the modern information society by enabling ubiquitous and diverse applications. 
From connecting humans to connecting intelligent agents \cite{zhang2025smart}, future sixth-generation (6G) networks \cite{dai2025metadetection} are anticipated to evolve into an integrated infrastructure beyond communication, tightly coupling AI and sensing to support emerging services \cite{choi2022smart}, including immersive communications, digital-twin interaction, and embodied-intelligence collaboration.
They are expected to operate across diverse space–air–ground–sea scenarios to broaden access coverage, extend across heterogeneous spectrum bands \cite{jornet2023above100ghz} and antenna scales to meet differentiated service demands, and integrate multimodal sensing information to enable the Synesthesia of Machines (SoM) \cite{cheng2024intelligent}.
Consequently, 6G systems will exhibit unprecedented heterogeneity across deployment scenarios, system configurations, and functionalities.
However, existing wireless physical-layer algorithms are largely model-driven, relying on propagation models, fixed configurations, and task-specific priors. 
{In practical heterogeneous systems, wireless algorithms often face unseen scenarios, diverse devices, and complex tasks. Hard-to-model non-idealities in these settings can induce severe model mismatch and substantially undermine algorithmic generalizability.}

With its powerful data-driven modeling capability and inherent robustness, AI is widely regarded as a key enabler for 6G to overcome the limitations of conventional algorithms.
It has been extensively explored for physical-layer tasks such as channel state information (CSI) estimation and prediction \cite{liu2024llm4cp,liu2025llm4wm}, and its demonstrated gains in spectral efficiency and sensing accuracy have drawn widespread interest.
For instance, companies such as NVIDIA have introduced AI-Radio Access Network (RAN) \cite{kundu2026ai-ran} to realize AI-native 6G networks.
Nevertheless, existing deep-learning-based wireless designs remain largely fragmented, with task-specific models tailored to individual physical-layer modules and system configurations, thereby limiting their broad applicability. 
In practice, a single base station (BS) may need to maintain hundreds of separately engineered models across diverse configurations and modules, leading to substantial deployment and management overhead. Moreover, these models are typically trained for specific scenarios and generalize poorly, so adapting to new environments often requires extensive labelled data and retraining, thereby impeding rapid deployment in dynamic settings. 
These limitations underscore the urgent need for a unified design framework for heterogeneous wireless systems.

Foundation models have reshaped the AI landscape \cite{abramson2024alphafold3,moor2023gmai,lu2025pretrained}, offering a compelling path to address these challenges. Pretrained with self-supervision on massive datasets, they adapt to diverse downstream tasks with minimal supervision and consistently outperform task-specific models. 
Given the central role of CSI in physical-layer design, a strong representation of CSI can benefit a wide range of channel-related communication and sensing tasks.
This motivates the development of a CSI-oriented wireless foundation model \cite{cheng2025som,liu2026wifo} that is pretrained at scale on wireless datasets to learn powerful representations transferable across scenarios, devices, and tasks. 
Such a general-purpose model could replace a large collection of separately designed task-specific models, substantially reducing or even eliminating the need for fine-tuning and thereby enabling a unified design paradigm.

Despite these promises, the development of wireless foundation models \cite{alikhani2024lwm,salihu2024ssiwloc,liu2025wifo,catak2025bert4mimo, zhao2024csibert2, jiang2025mimo} faces several fundamental challenges.
First, the construction of large-scale heterogeneous CSI datasets remains costly, while existing datasets are still limited in scale, source diversity, system heterogeneity, and openness, thus limiting the upper bound of model performance.
Second, heterogeneity in data dimensionality, distributions, and wireless task characteristics poses fundamental challenges to the network architecture and pretraining strategy design. 
{Prior studies have largely overlooked heterogeneous data and examined only a narrow set of wireless tasks.
For instance, LWM \cite{alikhani2024lwm} adopts masked channel modelling as a self-supervised pretraining strategy and is fine-tuned on only two types of tasks, including beam prediction. 
Our earlier work, WiFo \cite{liu2025wifo}, considers task-specific self-supervised pretraining and achieves superior zero-shot performance in time-domain and frequency-domain channel prediction.}
Third, unlike LLMs, wireless foundation models must deliver reliable outputs under the stringent inference-latency constraints imposed by real-time wireless systems, which have been largely overlooked in prior research.
More importantly, prior studies are limited to simulation-based validation, and the feasibility and performance advantages of deploying foundation models in real-world communication systems remain to be demonstrated.

In this work, we present WiFo-2, a versatile wireless foundation model that unifies the design of heterogeneous wireless systems. 
To support pretraining and evaluation, we construct the world’s largest open-source heterogeneous 3D space–time–frequency CSI dataset (LH-CSI), comprising 11.6 billion CSI points collected from 78 subsets. 
To effectively address task and data heterogeneity while substantially improving computational efficiency, WiFo-2 adopts
a masked denoising autoencoder (MDAE) architecture augmented with sparse mixture-of-experts networks. 
We further develop a two-stage pretraining strategy to learn generalizable wireless representations while enhancing channel reconstruction capability.

We demonstrate that WiFo-2 can be effectively applied to twelve wireless communication and sensing tasks in few-shot and even zero-shot settings.
For three channel reconstruction tasks, WiFo-2 surpasses the full-shot performance of task-specific models trained on 9,000 samples in zero-shot settings, while accurately estimating the confidence of reconstruction results.
More importantly, we reveal wireless scaling laws for WiFo-2 that are jointly shaped by data scale, heterogeneity, and model size, thereby validating its predictable scalability.
For nine CSI-related downstream tasks, WiFo-2 surpasses supervised approaches using only 1\% of the training samples, highlighting the superiority of its cross-task wireless representations.
In addition, we develop the first hardware prototype empowered by a wireless foundation model for over-the-air experiments, demonstrating its performance advantages across three practical communication and sensing tasks in real-world environments under stringent latency and reliability requirements.
Overall, through both simulation and real-world experiments, WiFo-2 demonstrates that wireless foundation models are a promising unified solution for heterogeneous wireless communication and sensing systems.
\section{RESULTS}
\subsection{Overview of WiFo-2}
\begin{figure*}[!htbp]
    \centering
    \makebox[\linewidth][r]{%
    \includegraphics[width=170mm]{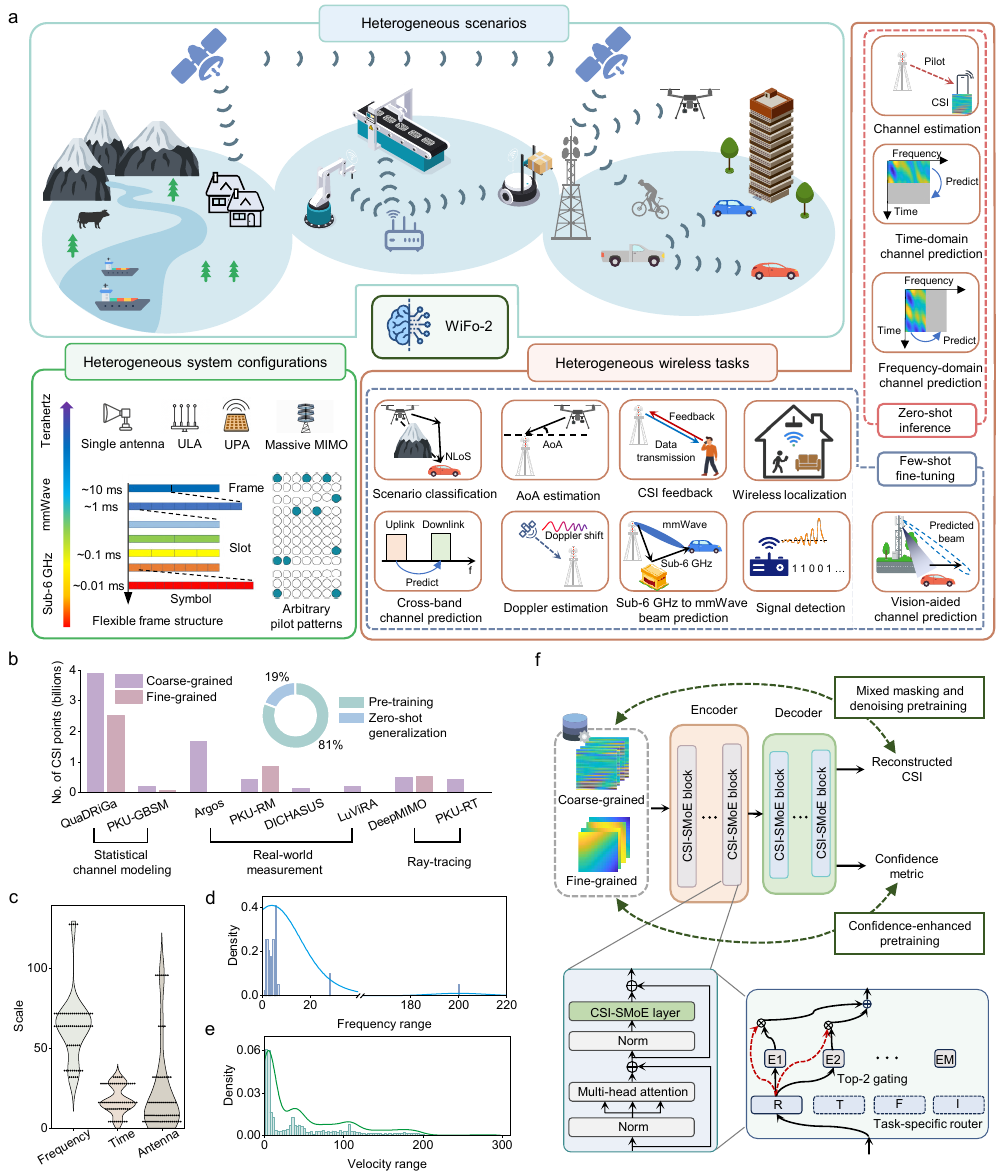}%
}
    \caption{Overview of WiFo-2. (a) WiFo-2 is a versatile foundation model for heterogeneous wireless systems, which generalizes across diverse scenarios, ranging from remote rural areas and smart factories to urban environments and non-terrestrial networks, accommodates heterogeneous system configurations including different frequency bands, antenna types, frame structures, and pilot patterns, and supports 12 CSI-related wireless communication and sensing tasks through few-shot fine-tuning or even zero-shot adaptation.
    (b) LH-CSI dataset composition and its split into pretraining and zero-shot generalization subsets.
    (c) Distribution of CSI size in LH-CSI across three dimensions.
    (d) Distribution of LH-CSI across frequency bands.
    (e) Distribution of LH-CSI across user velocities.
    (f) WiFo-2 is based on the MDAE architecture with CSI-SMoE layers and undergoes two-stage pretraining.}
    \label{fig1}
\end{figure*}
WiFo-2 is a versatile foundation model for heterogeneous wireless systems, which generalizes across diverse scenarios, accommodates various system configurations, and supports a wide range of CSI-related wireless communication and sensing tasks through few-shot fine-tuning or even zero-shot adaptation (Fig.~\ref{fig1}a).
For pretraining, we curated LH-CSI, an unprecedented large-scale heterogeneous space–time–frequency 3D CSI dataset, which contains 11.6 billion CSI points drawn from 8 distinct sources \cite{jaeckel2014QuaDRiGa, huang2023mixed, yaman2024luvira, dichasus2021, shepard2016argoschannel, alkhateeb2019deepmimo, sionna}, covering all three acquisition modalities, namely statistical channel modeling, real-world measurements, and ray-tracing (Fig.~\ref{fig1}b).
It comprises 78 coarse- and fine-grained sub-datasets that support tasks with diverse time–frequency sampling granularities, each exhibiting pronounced heterogeneity in its three-dimensional shape, including different numbers of subcarriers, antennas, and temporal sampling points (Fig. \ref{fig1}c).
In addition, LH-CSI spans frequencies from sub-6 GHz to terahertz, covers representative scenarios such as factories, unmanned aerial vehicles, and indoor and outdoor environments, and incorporates a wide range of user mobility speeds (Fig.~\ref{fig1}d-e).
LH-CSI was partitioned into an 81\% subset as pretraining sub-datasets and a 19\% subset reserved as zero-shot generalization sub-datasets for CSI reconstruction tasks.

{WiFo-2 adopts the proposed transformer-based MDAE architecture, which comprises a preprocessor for task-specific processing, an encoder for general CSI representation, and a decoder for CSI reconstruction (Fig. \ref{fig1}f).}
To enable unified learning across diverse CSI distributions and heterogeneous pretraining tasks, we introduce a CSI-SMoE layer for both the encoder and decoder, replacing the feedforward networks (FFNs) in standard transformer blocks. 
Specifically, each CSI-SMoE layer comprises multiple expert networks that are selectively activated, with the expert activation determined by a task-related gating mechanism conditioned on the input tokens (Note 12 of the Supplementary Information).
Moreover, this partial-activation mechanism preserves high model capacity while substantially reducing computational cost, enabling energy-efficient, low-latency communication.
WiFo-2 comprises five model variants of different scales. 
Unless otherwise specified, we use the large variant, which contains 100.86 million total parameters, 50.86 million of which are active.

WiFo-2 utilizes a two-phase pretraining strategy. 
The first phase comprises four mixed masking and denoising pretraining tasks, enabling it to acquire generalized and robust CSI representations.
Specifically, random, time, and frequency-masked reconstruction are conducted on the coarse-grained CSI dataset, while interpolation denoising is applied to the fine-grained CSI dataset. 
During pretraining, we randomize the prediction and downsampling ratios and inject Gaussian noise to improve WiFo-2’s robustness to varying reconstruction ratios and noise levels.
In the second phase, confidence-enhanced pretraining is introduced to enable confidence estimation for zero-shot channel reconstruction while retaining the capabilities learned in the first phase, which is crucial for reliability in real-world deployments.
After large-scale pretraining, WiFo-2 acquires generalizable CSI knowledge, supporting direct zero-shot transfer to three channel reconstruction tasks and few-shot adaptation to nine CSI-related tasks.

\subsection{WiFo-2 enables accurate and reliable zero-shot CSI reconstruction}
\begin{figure*}[!htbp]
    \centering
    \includegraphics[width=1\linewidth]{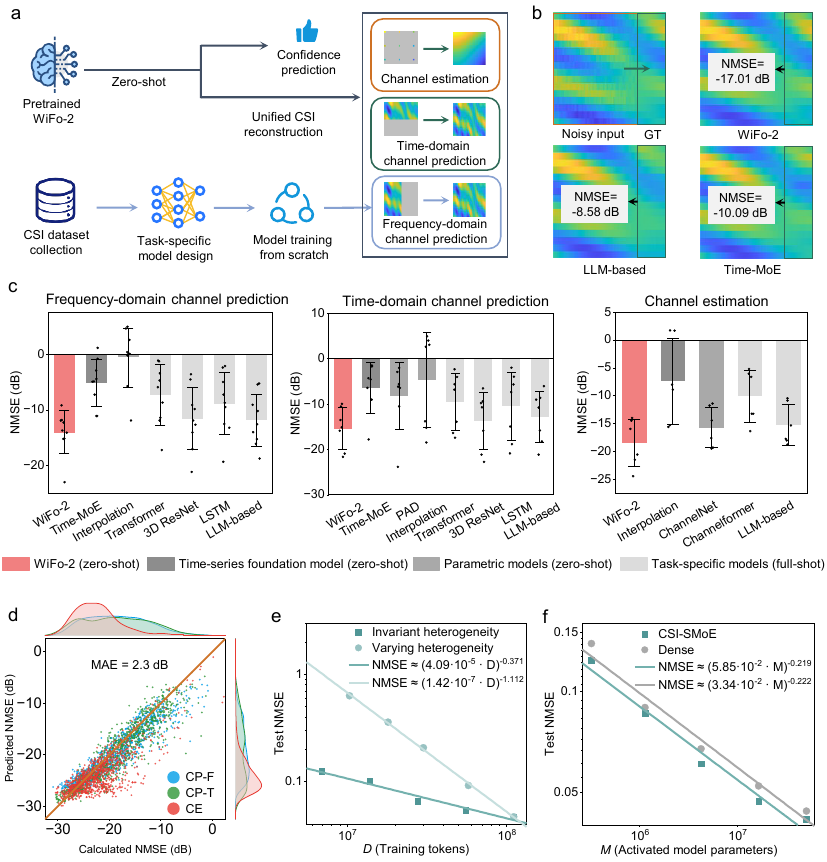}
    \caption{Performance analysis of WiFo-2 on zero-shot channel reconstruction tasks. 
    (a) Workflow comparison between WiFo-2 and task-specific models for channel reconstruction tasks.
    (b) Visual comparison of CSI for frequency-domain channel prediction.
    {(c) Comparison of channel reconstruction accuracy between WiFo-2 and baseline methods at a low reconstruction ratio and an SNR of 10 dB.
    Error bars indicate $\pm$ standard deviation (SD) calculated across all zero-shot datasets, and dots denote individual dataset results.}
    (d) Visualization of WiFo-2 confidence prediction performance across three channel reconstruction tasks.
    {(e) Scaling analysis of zero-shot channel reconstruction performance with respect to training tokens. The $R^2$ values for the invariant-heterogeneity and varying-heterogeneity are 0.9682 and 0.9987, respectively.
    (f) Scaling analysis of zero-shot channel reconstruction performance with respect to activated model parameters.
    The $R^2$ values for the CSI-SMoE version and dense version are 0.9756 and 0.9746, respectively.}}
    \label{fig2}
\end{figure*}
Channel reconstruction is a fundamental task in wireless system design, and its accuracy directly determines the upper performance bound of CSI-related tasks. It aims to recover the full CSI from partial pilot observations, including time-domain and frequency-domain channel prediction, and channel estimation. Existing task-specific AI models are typically designed for particular channel reconstruction tasks and system configurations, but generalize poorly to new scenarios and often require additional data collection and model retraining. In contrast, WiFo-2 unifies zero-shot channel reconstruction across scenarios, system configurations, and tasks within a single model, while additionally providing confidence predictions for the reconstructed results (Fig.~\ref{fig2}a).

We evaluated WiFo-2 on the zero-shot generalization split of the LH-CSI dataset, which includes nine coarse-grained CSI sub-datasets for channel prediction and six fine-grained sub-datasets for channel estimation. 
{To evaluate robustness, we tested two SNR settings (10 dB and 20 dB), reporting performance across both low and high reconstruction ratios.
For the channel prediction task, the low and high prediction ratios were set to 25\% and 50\%, respectively. 
For the channel estimation task, the low and high estimation ratios corresponded to pilot-symbol placement proportions of $(1/4, 1, 1/12)$ and $(1/8, 1, 1/24)$ along the time, antenna, and subcarrier dimensions, respectively.}
We compared three categories of methods, namely supervised AI models (Transformer \cite{jiang2022accurate}, 3D ResNet \cite{feichtenhofer2019slowfast}, ChannelNet \cite{soltani2019deep}, Channelformer \cite{luan2023channelformer}, LSTM-based schemes \cite{jiang2020deep}, and LLM-based approaches \cite{liu2024llm4cp}), parametric models (interpolation and PAD\cite{yin2020curse}), and Time-MoE \cite{xiaoming2025time} as a universal time-series foundation model. 
Both WiFo-2 and Time-MoE are utilized without any fine-tuning, whereas all task-specific AI models and the LLM-based scheme are trained separately for each dataset and reconstruction ratio.

Figure~\ref{fig2}c compares different methods at a low reconstruction ratio and an SNR of 10 dB. WiFo-2 achieves state-of-the-art zero-shot performance, even surpassing all task-specific models trained with full samples. 
Specifically, for channel estimation, time-domain channel prediction and frequency-domain channel prediction, WiFo-2 outperforms the second-best method by 2.82 dB, 1.63 dB and 0.26 dB in NMSE, respectively. 
{Even on challenging real-world measurement datasets, including Argos and PKU-RM, WiFo-2 outperforms Transformer and LSTM in both frequency-domain and time-domain channel prediction, while achieving SOTA channel estimation performance with a 1.58 dB gain over the second-best method.}
Figure~\ref{fig2}b provides a visual comparison for frequency-domain channel prediction, showing that WiFo-2 reconstructs clearer and more accurate textures than the competing methods. 
Further experimental results confirm that WiFo-2 achieves the lowest NMSE across all reconstruction ratios and SNR levels for all three tasks (Note 9 of the Supplementary Information). These results demonstrate its general and robust zero-shot channel reconstruction capability, as well as strong generalization to unseen scenarios and system configurations.

To validate the effectiveness of WiFo-2’s confidence prediction, we assess the accuracy of the predicted NMSE on zero-shot generalization datasets. 
Figure \ref{fig2}d presents the scatter plot and distribution of the actual reconstruction NMSE and the predicted confidence. 
The prediction error remains low overall, with a mean absolute error (MAE) of only 2.3 dB.
{These results indicate that WiFo-2 provides accurate confidence estimates that can inform model updates and communication-strategy selection, thereby substantially enhancing reliability and adaptability in real-world deployments. 
For instance, when the current model output has low confidence, the system can switch to a more conservative conventional algorithm or fine-tune the model to maintain the reliability of the communication link. 
In addition, the confidence output can be used to predict the channel estimation accuracy under different candidate pilot densities, based on which the system can select a pilot configuration that balances pilot overhead and retransmission risk \cite{cheng2026large}. 
This improves the overall communication throughput without requiring additional channel-quality measurements.}

Furthermore, we analyze the scaling laws of zero-shot channel reconstruction performance with respect to both data and model scale (Fig. \ref{fig2}e and f). 
For data scaling, we adopt the base version of WiFo-2 and consider two settings: one with fixed heterogeneity, where the number of pretraining subdatasets is kept constant, and only the sampling ratio within each subdataset is reduced, and the other with varying heterogeneity, where the number of subdatasets is reduced. 
In both cases, WiFo-2 follows a clear scaling law with training tokens. 
Moreover, increasing data heterogeneity markedly improves zero-shot performance. 
{At the same $10^7$ training tokens, the high-heterogeneity dataset reduces zero-shot NMSE by 8.0 dB relative to its low-heterogeneity counterpart, highlighting the decisive role of heterogeneous data in enhancing the generalization of wireless foundation models. 
It can be explained by the fact that heterogeneous data force WiFo-2 to learn the intrinsic space-time-frequency structure of CSI across different scenarios and system configurations, rather than memorizing statistical shortcuts tied to a specific dataset. 
As a result, WiFo-2 achieves better generalization to unseen data distributions.}
For model scaling, we compare WiFo-2 with a dense variant in which CSI-SMoE layers are replaced by feed-forward networks with the same number of active parameters. 
To reach the same NMSE, the CSI-SMoE design requires about 33\% fewer active parameters, substantially reducing computation and inference latency for deployment on resource-constrained terminals. 
These scaling laws provide practical guidance for both model size selection and dataset construction.

\subsection{WiFo-2 facilitates CSI-related downstream tasks with few-shot fine-tuning}
\begin{figure*}[!t]
    \centering
    \makebox[\linewidth][r]{%
    \includegraphics[width=170mm]{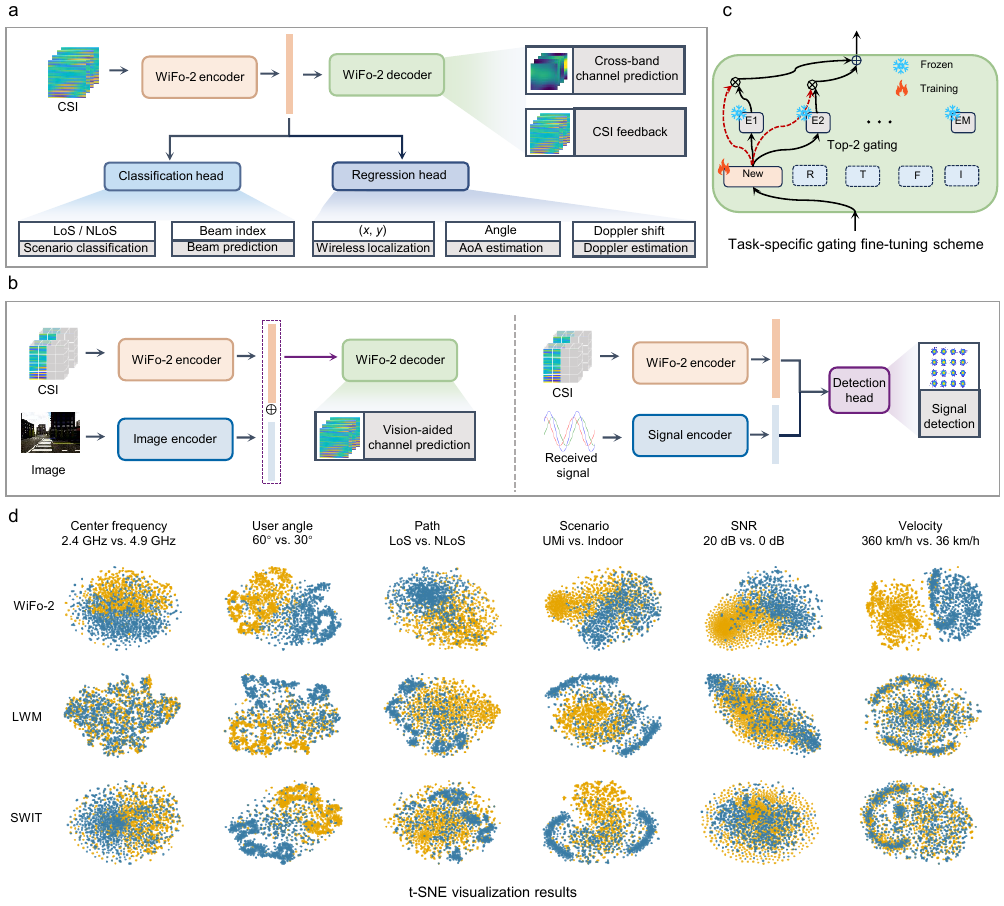}%
}
    \caption{Fine-tuning WiFo-2 on CSI-related downstream tasks. 
    (a) Fine-tuning pipelines of WiFo-2 for CSI-only wireless tasks. 
    (b) Fine-tuning pipelines of WiFo-2 for multi-modal wireless tasks. 
    (c) An illustration of the proposed task-specific gating fine-tuning scheme.
    (d) The t-SNE visualization of final-layer encoder tokens for WiFo-2, LWM, and SWIT across 6 datasets.}
    \label{fig3}
\end{figure*}
\begin{figure*}[!t]
    \centering
    \makebox[\linewidth][r]{%
    \includegraphics[width=170mm]{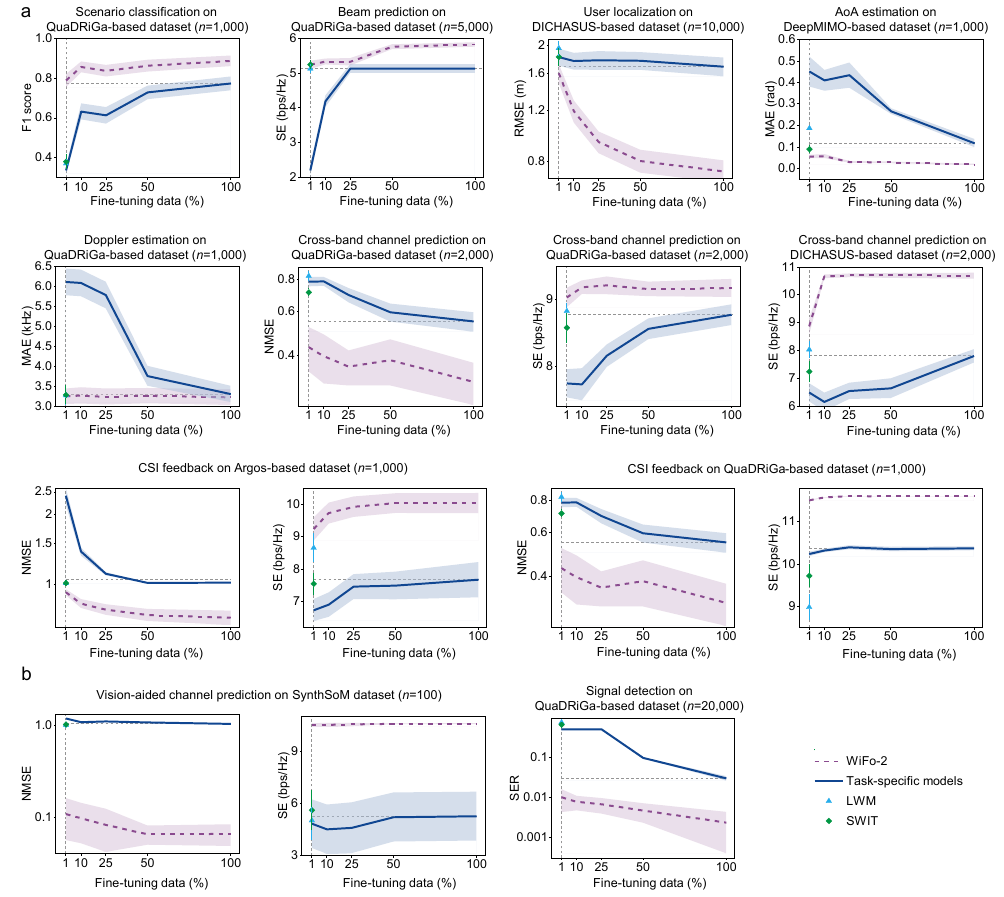}%
}
    \caption{Results of WiFo-2 on CSI-related downstream tasks under different fine-tuning data ratios. 
    {$n$ denotes the number of training samples.
    The shaded area represents the 95\% confidence interval (CI).}
    {Higher values are better for F1 score and SE, whereas lower values are better for RMSE, MAE, NMSE, and SER.}
    (a) Performance comparisons on CSI-only wireless tasks.
    (b) Performance comparisons on multi-modal wireless tasks.
    }
    \label{fig4}
\end{figure*}
Beyond channel reconstruction, we further evaluate WiFo-2 on nine wireless communication and sensing tasks via few-shot fine-tuning, including tasks based on CSI alone and those integrating additional modalities. 
For different tasks, we design tailored fine-tuning schemes that leverage the respective strengths of the encoder and decoder.
{In general, the encoder of WiFo-2 is used for CSI feature extraction, while the decoder is used for CSI reconstruction. 
For multi-modal wireless tasks, a new modality is processed by a dedicated modality-specific encoder and then concatenated with the CSI features for subsequent processing. 
In addition, we propose a task-specific gating fine-tuning scheme, in which a new router dedicated to each task is added to the CSI-SMoE layer of the WiFo-2 encoder during fine-tuning (Fig. \ref{fig3}c). 
For the WiFo-2 encoder, only the router parameters are updated, while the expert weights remain frozen. This enables effective transfer of general CSI knowledge across eight CSI-related downstream tasks.}
Compared with existing gating networks, the proposed gating fine-tuning scheme achieves state-of-the-art performance across all tasks while introducing negligible fine-tuning overhead (Note 14 of the Supplementary Information). 

For each task, we curated datasets either derived from publicly available CSI datasets or constructed via simulation, spanning diverse data sources.
We adopted both pretrained wireless foundation models (including LWM \cite{alikhani2024lwm} and SWIT \cite{salihu2024ssiwloc}) and SOTA task-specific AI models for comparison.
Each wireless foundation model is fine-tuned on the constructed datasets in a few-shot manner with its backbone kept frozen, whereas task-specific models are trained from scratch.
For each task, we report model performance under different fine-tuning data ratios. Across all tasks and datasets, WiFo-2 surpasses the full-shot performance of task-specific models using only 1\% of the fine-tuning samples, while also consistently outperforming LWM and SWIT under fine-tuning. 
Detailed analyses are provided below, and the implementation details are illustrated in Note 7 of the Supplementary Information. 

\textbf{Scenario classification.}
The task aims to infer from CSI samples whether a direct LoS path is present in the wireless link.
We construct a dataset based on QuaDRiGa and use the F1 score to evaluate LoS/NLoS classification accuracy. 
We adopt ST-CNN \cite{sun2022channel} as the SOTA task-specific AI baseline. 
Even with 99\% less fine-tuning data, WiFo-2 surpasses ST-CNN by 1.87\%, evidencing its stronger discrimination of multipath components.

\textbf{Sub-6 GHz to mmWave beam prediction.}
For a dual-band communication system, this task aims to predict the optimal mmWave beam index according to the CSI of the sub-6 GHz frequency band.
We construct a QuaDRiGa-based dataset and evaluate performance using top-1 classification accuracy and spectral efficiency (SE), the latter obtained via link-level simulations. 
AI models referenced in \cite{alrabeiah2020deep} are used as task-specific baselines for comparison. 
WiFo-2 exceeds the task-specific model by $1.87\%$ in accuracy and $2.08\%$ in spectral efficiency with 1\% fine-tuning data, demonstrating its superior spatial feature extraction capability.

\textbf{Wireless localization.}
The task aims to perform two-dimensional user localization from CSI. 
We construct a wireless localization dataset based on the open-source real-measurement dataset DICHASUS and use root-mean-square error (RMSE) to quantify localization performance. 
WiT \cite{salihu2022attention} serves as the SOTA task-specific baseline. 
Even with 99\% less fine-tuning data, WiFo-2 achieves a 5.04\% lower RMSE than WiT, highlighting its ability to extract localization-relevant features directly from raw CSI.

\textbf{AoA estimation.} 
The task aims to estimate the AoA of the dominant path between the transmitter and the receiver from CSI. 
We construct a ray-tracing CSI dataset based on DeepMIMO and utilize the mean absolute error (MAE) of the estimated angle as the evaluation metric. 
We build a task-specific AI model based on ResNet similar to \cite{pan2025large}.
Remarkably, with only 1\% of the fine-tuning data, WiFo-2 attains an exceptionally low MAE of 0.055 rad, reducing the error by 52.97\% compared with the full-shot ResNet baseline, which validates its ability to capture dominant path information from CSI.

\textbf{Doppler estimation.} 
This task is designed to extract Doppler shift information from CSI. 
We construct a wireless dataset for satellite communication scenarios using the QuaDRiGa simulator and utilize the MAE of Doppler shift estimation as the metric.
We adopted a CNN-Transformer style solution \cite{ahmadi2025semanticaware} as the task-specific baseline.
WiFo-2 achieves a 1.78\% reduction in MAE while only using 1\% of the fine-tuning data, highlighting its strong capability to capture the dominant temporal dynamics of wireless channels.

\textbf{Cross-band channel prediction}.
The task aims to infer the CSI at a nonadjacent frequency based on the CSI from another band.
We construct two datasets for evaluation, one generated by QuaDRiGa simulations and another built from the DICHASUS dataset, and we evaluate model performance using both the NMSE of the prediction and the system SE. 
RF-Diffusion\cite{chi2024rf}, a powerful diffusion-based generative model for RF data synthesis, was utilized as the task-specific baseline.
{On both datasets, WiFo-2 reduces NMSE by at least 0.14 dB and increases SE by at least 2.97\% relative to the fully supervised model with 1\% fine-tuning data, which validates its strong capability for extrapolation across frequency.}

\textbf{CSI feedback.}
Typical CSI feedback includes compressing CSI at the transmitter and reconstructing the original CSI at the receiver. 
This task focuses on post-processing the CSI reconstructed by the classical CSI feedback algorithm \cite{wen2018deep} to further improve reconstruction fidelity. 
We construct two datasets: one based on QuaDRiGa simulations and another derived from the open-source Argos measurement dataset.
We evaluate the reconstruction performance via NMSE of the reconstructed CSI and the system SE. 
We adopt TransNet \cite{cui2022transnet} as a competitive task-specific baseline. 
{WiFo-2 reduces NMSE by at least 4.17 dB and increases SE by at least $12.44\%$ relative to fully supervised solutions while reducing the amount of fine-tuning data by 99\%, which validates its strong denoising ability.}

\textbf{Vision-aided frequency-domain channel prediction.}
The task aims to enhance frequency-domain channel prediction by exploiting environmental images captured by co-located cameras. 
We construct a dataset based on the open-source multi-modal sensing-communication dataset SynthSoM \cite{cheng2025synthsom}, and we evaluate models with the prediction NMSE and the system SE. 
We construct a task-specific baseline using a vision-CSI fusion network similar to that in \cite{nam2025mmvr_csi}.
In this highly challenging task, LWM and SWIT perform poorly, whereas WiFo-2 markedly outperforms the task-specific model even with 99\% less fine-tuning data. 
{WiFo-2 achieves a 10.37 dB reduction in NMSE and a 99.61\% gain in SE, indicating that its learned general representations are more readily aligned with other modalities and thus better support cross-modal generalization and enhancement.}

\textbf{Signal detection}.
The task aims to recover the transmitted symbols from the received symbols using the estimated CSI. 
We construct a simulation dataset with QuaDRiGa and evaluate performance using the symbol error rate (SER) of the demodulated symbols. 
We adopt OAMPNet \cite{he2020model} as the SOTA task-specific baseline for the signal detection.
Relative to the task-specific model, WiFo-2 reduces SER by 66.10\% while using only 1\% of the fine-tuning data, a regime in which LWM and SWIT are largely unable to operate effectively. 
This again demonstrates that universal wireless representations possess strong cross-modal generalization capabilities in extremely low-data settings.

To intuitively illustrate the wireless knowledge captured by the general representations of WiFo-2, we apply t-SNE \cite{vandermaaten2008tsne} to compare the token embeddings from the final encoder layer of WiFo-2 with those derived from LWM and SWIT.
We construct 6 QuaDRiGa-based simulation datasets for controlled comparisons. 
In each set, only one factor varies, namely center frequency, user angle, the presence or absence of a LoS path, scenario category, SNR, or user velocity. 
Across all sets, WiFo-2 yields more clearly separated clusters, whereas the baselines show mixed and poorly separated patterns. 
These results intuitively demonstrate the strong representational capacity of WiFo-2 for heterogeneous CSI distributions.

\subsection{Experimental validation with a hardware prototype}
\begin{figure*}[!t]
    \centering
    \includegraphics[width=1\linewidth]{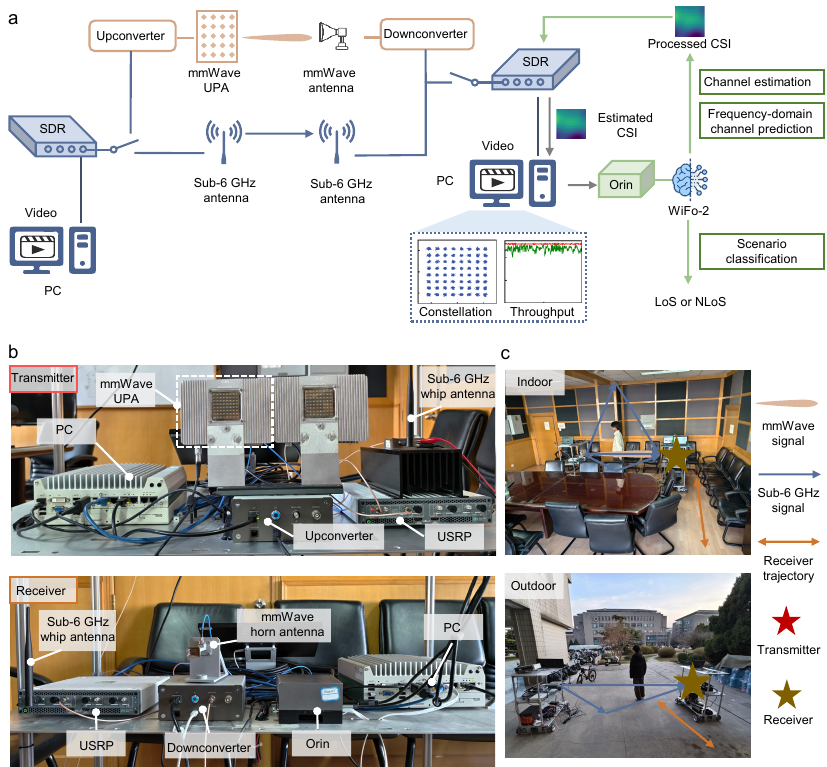}
    \caption{Overview of the experimental validation. (a) Overall workflow of the hardware prototype. (b) Photographs of the principal hardware components. (c) Test environments, including indoor and outdoor scenarios.}
    \label{fig5}
\end{figure*}
\begin{figure*}[!t]
    \centering
    \includegraphics[width=1\linewidth]{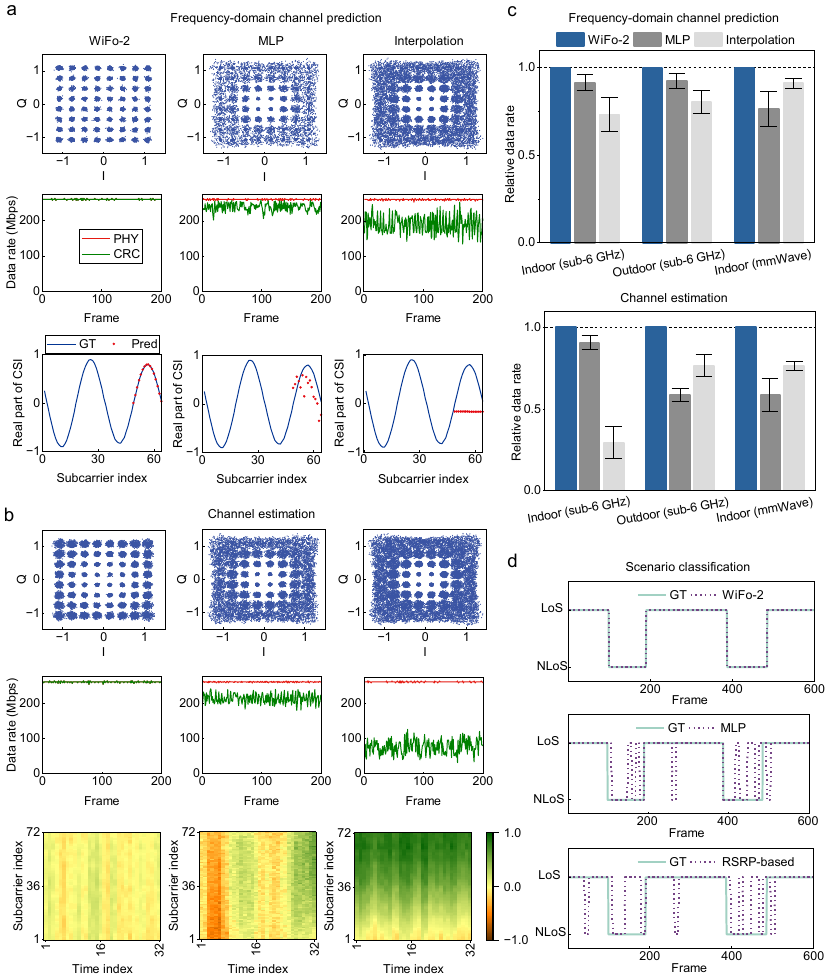}
    \caption{Over-the-air test results. (a) Channel prediction results under the indoor and sub-6 GHz conditions, including receiver constellation diagrams, data rates, and CSI prediction visualizations. (b) Channel estimation results under the indoor and sub-6 GHz conditions, including receiver constellation diagrams, data rates, and CSI estimation errors. {(c) Relative data rates of different methods for the two tasks across three scenarios, measured as the ratio of CRC-passed frames to total frames. Error bars indicate $\pm$ SD calculated across all frames.} (d) Performance comparisons of different methods on scenario classification.}
    \label{fig6}
\end{figure*}
To validate the practical deployability of WiFo-2 and quantify its performance gains in real-world systems, we developed a wireless foundation model-empowered hardware prototype to evaluate it across three representative wireless communication and sensing tasks: frequency-domain channel prediction, channel estimation, and scenario classification.
As illustrated in Fig.~\ref{fig5}a, the deployment prototype consists of a transmitter and a receiver built on software-defined radios (Ettus USRP X410). 
The system operates in the sub-6 GHz band and can be extended to the mmWave band via external frequency converters. 
Following the 5G NR frame structure, it supports video transmission, and Demodulation Reference Signal (DM-RS) pilot symbols are inserted into selected time-frequency resources for channel estimation. 
At the receiver, the estimated CSI is extracted and transmitted through a wired TCP link to an NVIDIA edge platform (Jetson AGX Orin) for processing, after which the processed CSI is written back to replace the original estimates for subsequent demodulation. 
The receiver also provides real-time displays of throughput, constellation diagrams, and reconstructed video to showcase the transmission performance.

To evaluate the model’s robustness across different environments, we considered both indoor and outdoor scenarios.
As shown in Fig.~\ref{fig5}c, the indoor scenario features a rich scattering environment comprising walls, tables, chairs, and moving pedestrians, and was evaluated at 4.0 GHz and 28.0 GHz. 
By contrast, the outdoor scenario is more open with fewer scatterers and was evaluated only at 4.0 GHz. 
In both cases, the transmitter was fixed, whereas the receiver moved linearly at approximately $1\,\mathrm{m/s}$. 
Pedestrians moved slowly between them, frequently blocking the LoS path.

The system has an operational bandwidth of 97.6 MHz, covering 3,252 subcarriers. 
For each task, the data are processed in units of ten consecutive frames of the time–frequency resource grid, with each unit spanning all subcarriers in frequency and 200 slots in time (100 ms). 
For the channel prediction task, the resource grid is partitioned into basic prediction blocks, each covering 64 subcarriers with a subcarrier spacing of 120 kHz and 32 time samples at intervals of 2.5 ms. 
Using the CSI of the first 48 subcarriers, we predict the CSI of the remaining 16 subcarriers, thereby reducing the pilot overhead by 25\%. 
For the channel estimation task, the grid is partitioned into basic estimation blocks, each covering 72 subcarriers with a spacing of 30 kHz and 32 time samples at intervals of 2.5 ms. 
Sparse-pilot channel estimation is performed with pilot densities of 1/8 in time and 1/24 in frequency, reducing the pilot overhead by 99.5\%.
For the scenario classification task, we determine the scenario type based on the CSI of basic estimation blocks.

{We adopt an MLP as the task-specific baseline owing to its suitability for edge deployment, and specifically trained a dense-architecture WiFo-2 model comprising a four-layer encoder and a two-layer decoder, with a feature dimension of 512 and 13.29 M parameters.}
We then collected 1,000 CSI samples from the testing scenario to fine-tune WiFo-2 and train the MLP model.
{To satisfy the real-time requirement of 100 ms, we apply INT8 quantization to WiFo-2 using NVIDIA TensorRT, reducing the full-operation inference latency for channel prediction and channel estimation to 31.94 ms and 21.48 ms, respectively.}
For the indoor scenario at 4.0 GHz, the performance of the different models on the frequency-domain channel prediction and channel estimation tasks is shown in Fig.~\ref{fig6}a and b, respectively, including receiver-side constellation diagrams, the physical-layer data rate together with the cyclic redundancy check (CRC)-verified data rate, and CSI visualizations. 
{Over the 200 consecutive frames observed, corresponding to 2,000 ms, WiFo-2 consistently produces tightly clustered constellation points, enables stable error-free transmission, and yields highly accurate CSI prediction and estimation.}
By contrast, the MLP- and interpolation-based methods incur larger errors, leading to constellation dispersion and throughput degradation.
We further compare performance under indoor 28.0 GHz and outdoor 4.0 GHz conditions, as shown in Fig.~\ref{fig6}c. 
{WiFo-2 consistently achieves error-free transmission, whereas the baseline methods suffer varying degrees of throughput degradation. 
These results demonstrate that, with only few-shot fine-tuning, WiFo-2 can substantially reduce pilot overhead across diverse real-world communication scenarios without compromising throughput, even in complex outdoor environments and previously unseen frequency bands.}
For the scenario classification task, we consider the indoor scenario at 4.0 GHz and additionally compare against a reference signal received power (RSRP)-based method that makes decisions based on a fixed power threshold. 
Compared with the baseline methods, only WiFo-2 can reliably identify the scenario type over long time horizons, highlighting its superior capability to extract environmental features from CSI.

\section{DISCUSSION}
In this study, we introduce WiFo-2, a wireless foundation model for the general-purpose design of wireless communication and sensing systems. 
WiFo-2 is designed to address the fundamental heterogeneity of 6G wireless environments by enabling compatibility across diverse system configurations while supporting ubiquitous coverage and enhanced sensing. 
{Compared with existing studies on wireless foundation models, WiFo-2 represents a substantial advancement in terms of the dataset scale and diversity, the number of supported tasks, generalization performance, and hardware prototype implementation (Note 1 of the Supplementary Information).}
Through a dedicated architecture and a two-stage pretraining strategy on an unprecedentedly large-scale heterogeneous LH-CSI dataset, WiFo-2 acquires general wireless knowledge that transfers effectively across 12 communication and sensing tasks. 
For channel reconstruction, its zero-shot performance surpasses the task-specific models trained on 9,000 samples, while also providing accurate confidence estimates for reconstructed outputs to improve transmission reliability. 
On downstream tasks, WiFo-2 remarkably exceeds the full-shot performance of task-specific baselines using only 1\% of the fine-tuning data. 
More importantly, we observe a distinctive scaling law in the wireless domain, in which performance depends not only on model size and dataset scale, but also strongly on dataset heterogeneity. 
We further develop, to our knowledge, the first wireless foundation model-empowered hardware prototype, and verify its advantages across multiple frequency bands, diverse scenarios, and three wireless communication and sensing tasks. 
Collectively, these results support WiFo-2 as a unified paradigm for wireless system design, overcoming the limitations of isolated task-specific models while significantly reducing the number of required models and the cost of fine-tuning.

Our work extends the foundation-model paradigm beyond language, vision, and biological sequences to a new physical modality—wireless channels—which constitute the most fundamental and generalizable representations across diverse environments, system configurations, and wireless tasks. 
Therefore, our approach substantially simplifies system design and opens a scalable route towards improved performance. 
In particular, scaling model capacity and expanding the size and diversity of pretraining data can enhance the model’s understanding of wireless channels, with broad downstream benefits for a wide range of CSI-related wireless tasks during deployment.
Its successful deployment in real-world engineered wireless systems under stringent storage, computational, and latency constraints further underscores its practical viability.

Despite these advances, research on wireless foundation models still faces several open issues. 
On the data side, the performance of WiFo-2 is more constrained by the scale and diversity of the data than by the model size.
Although we have built an unprecedentedly large and heterogeneous CSI dataset, it remains dominated by simulated data and is largely confined to sub-6 GHz bands. 
Therefore, it is essential for industrial and academic partners in wireless communications to jointly develop large-scale, open-source, measurement-based CSI datasets, thereby fundamentally lifting the performance ceiling of wireless foundation models.
{On the hardware deployment side, the stringent latency and power constraints of practical wireless systems should be further considered. 
The developed hardware prototype is still preliminary, as it uses the edge-AI platform as the inference device and relaxes the inference-time requirement to 100 ms by jointly processing every 10 frames.
However, the additional 100 ms latency remains incompatible with the millisecond-level latency requirements of practical wireless systems. 
Moreover, the approximately 5 W power overhead is still considerable for BSs and prohibitive for mobile phones. 
These limitations call for lightweight architectures and deployment-acceleration techniques for wireless foundation models to enable practical adoption.}

In summary, this work presents a scalable foundation model that enables a unified approach to wireless system design across diverse environments, system configurations, and tasks, offering a practical and promising path towards AI-native 6G wireless systems.
\section{METHODS}
\subsection{Large-scale dataset construction}
To pretrain WiFo-2 and evaluate its zero-shot generalization performance on channel-reconstruction tasks, we curated LH-CSI, a large-scale heterogeneous three-dimensional CSI dataset comprising 11.6 billion CSI points. 
The dataset integrates data from eight distinct sources, including QuaDRiGa \cite{jaeckel2014QuaDRiGa}, PKU-GBSM \cite{huang2023mixed}, LuViRA \cite{yaman2024luvira}, DICHASUS \cite{dichasus2021}, Argos \cite{shepard2016argoschannel}, PKU-RM, DeepMIMO \cite{alkhateeb2019deepmimo}, and PKU-RT \cite{sionna}, covering all three CSI acquisition modalities of statistical channel modelling, real-world measurements, and ray tracing.
Each source dataset comprises several sub-datasets of either coarse-grained or fine-grained types, with 9,000, 1,000, and 2,000 CSI samples designated for training, validation, and testing, respectively.
The dataset covers a broad range of wireless communication scenarios, including factories, schools, and unmanned aerial vehicle (UAV) communications. It features rich system configurations with carrier frequencies spanning from sub-6 GHz to the terahertz band and a diverse spectrum of user-mobility speeds. 
Detailed descriptions of each source and the system configurations are given in Note 3 of the Supplementary Information.

\subsection{Model architecture}
Building on the strong performance of the MAE \cite{he2022mae} design, we propose an MDAE as the network architecture of WiFo-2, in order to accommodate masked reconstruction and denoising. 
The architecture consists of three modules: a preprocessor that converts raw CSI into tokens through task-specific processing, an encoder that transforms these tokens to capture general CSI representations, and a decoder that reconstructs the CSI from the encoded features.
Further details about the model architecture 
are provided in Note 4 of the Supplementary Information. 

\subsection{Pretraining scheme and applications}
We adopted a two-phase pretraining strategy. In the first phase, the model is trained on four tasks—random-masked reconstruction, time-masked reconstruction, frequency-masked reconstruction, and interpolation denoising—using reconstruction MSE as the objective. In the second phase, termed confidence-enhanced pretraining, the model is trained to predict the normalized mean squared error (NMSE) of channel reconstruction, with the MSE of the confidence estimate as the loss. After this two-stage pretraining, WiFo-2 can be directly applied to CSI reconstruction tasks using a task-specific gating scheme. Additional details of pretraining and fine-tuning are provided in Note 5 of the Supplementary Information.
\section{DATA AND CODE AVAILABILITY}
Datasets, pretrained model weights, and source code associated with this work are publicly available in the GitHub repository at https://github.com/PKU-PCNI/WiFo-2.
\section{FUNDING}
This work was supported by the National Natural Science Foundation of China (62125101, 62401488, 62571007 and U23A20339), the New Cornerstone Science Foundation through the Xplorer Prize, the Guangzhou–The Hong Kong University of Science and Technology (Guangzhou) Joint Funding Scheme (2025A03J3878), the Guangdong Provincial Key Laboratory of Integrated Communication, Sensing and Computation for Ubiquitous Internet of Things (2023B1212010007), the Guangzhou Municipal Science and Technology Project (2024D03J0008), the Guangdong Provincial Project (2023ZT10X009) and the Fundamental Research Funds for the Central Universities, Peking University.
\section{AUTHOR CONTRIBUTIONS}
X.C. conceived the idea and planned the study. 
B.X.L. developed the methodology, performed model pretraining, constructed the dataset, and wrote the original draft. 
X.Y.L. fine-tuned the model for downstream tasks and contributed to the dataset construction. 
B.X.L., X.Y.L., and X.S.C. built the hardware prototype.
S.J.G. contributed to the methodology. 
S.J.G., X.S.C., X.C., and L.Q.Y. supervised the study, reviewed, and edited the manuscript. 
All authors contributed to the writing and revision of the manuscript and approved the final version.

\textbf{\textit{Conflict of interest statement.}} None declared. 
\bibliographystyle{nsr}
\bibliography{nsr_sample}
\end{document}